\documentclass[preprint2]{aastex}

\shorttitle{UCAC3 Proper Motion Survey}
\shortauthors{Finch}

\begin{document}

\title{UCAC3 Proper Motion Survey. I. \\ DISCOVERY OF NEW PROPER
  MOTION STARS IN UCAC3 \\ WITH 0$\farcs$40 yr$^{-1}$ $>$ $\mu$ $\ge$
  0$\farcs$18 yr$^{-1}$ BETWEEN DECLINATIONS $-$90$\degr$ and
  $-$47$\degr$}

\author{Charlie T. Finch, Norbert Zacharias}

\email{finch@usno.navy.mil}

\affil{U.S. Naval Observatory, Washington DC 20392--5420}

\author{Todd J. Henry}

\affil{Georgia State University, Atlanta, GA 30302--4106}


\begin{abstract}

Presented here are 442 new proper motion stellar systems in the
southern sky between declinations $-$90$\degr$ and $-$47$\degr$ with
0$\farcs$40 yr$^{-1}$ $>$ $\mu$ $\ge$ 0$\farcs$18 yr$^{-1}$.  These
systems constitute a 25.3\% increase in new systems for the same
region of the sky covered by previous SuperCOSMOS RECONS (SCR)
searches that used Schmidt plates as the primary source of discovery.
Among the new systems are 25 multiples, plus an additional seven new
common proper motion companions found to previously known primaries.
All stars have been discovered using the third U.S. Naval Observatory
(USNO) CCD Astrograph Catalog (UCAC3).  A comparison of the UCAC3
proper motions to those from the Hipparcos, Tycho-2, Southern Proper
Motion (SPM4), and SuperCOSMOS efforts is presented, and shows that
UCAC3 provides similar values and precision to the first three
surveys.  The comparison between UCAC3 and SuperCOSMOS indicates that
proper motions in RA are systematically shifted in the SuperCOSMOS
data but are consistent in DEC data, while overall showing a
significantly higher scatter.  Distance estimates are derived for
stars having SuperCOSMOS Sky Survey (SSS) $B_J$, $R_{59F}$, and
$I_{IVN}$ plate magnitudes and Two-Micron All Sky Survey (2MASS)
infrared photometry.  We find 15 systems estimated to be within 25 pc,
including UPM 1710-5300 our closest new discovery estimated at 13.5
pc.  Such new discoveries suggest that more nearby stars are yet to be
found in these slower proper motion regimes, indicating that more work
is needed to develop a complete map of the solar neighborhood.

\end{abstract}

\keywords{solar neighborhood --- stars: distances --- stars:
statistics --- surveys --- astrometry} 

\section{INTRODUCTION}

Proper motion surveys were once in the forefront of astronomy
research.  Providing a vast library of low mass stars, proper motion
surveys provide a wealth of information to astronomers studying
stellar populations, and in particular the stellar luminosity and mass
functions that reveal how the Galaxy's stellar mass is divided among
different types of stars.  Proper motion surveys of faint objects,
such as red dwarfs, subdwarfs, and white dwarfs, play a crucial role
in identifying the Sun's nearest neighbors.

Historically, proper motion studies have been carried out by blinking
photographic plates taken at different epochs to detect the changing
positions of stars.  The pioneering surveys of the second half of the
last century include the Lowell Proper Motion Survey
\citep{1971lpms.book.....G,1978LowOB...8...89G}, the Luyten
Half-Second catalog (LHS) \citep{1979lccs.book.....L} and the New
Luyten Two-Tenths catalog (NLTT) \citep{1980PMMin..55....1L}.  With
the aid of plate scanning machines and high powered computers, the
traditional techniques used for proper motion studies are carried out
in much the same way, only now using digitized images of the
photographic plates.  Utilizing various techniques and plate sets,
these new computerized searches have revealed many new proper motion
systems.  Recent surveys of the southern sky --- the region targeted
by the present effort --- include \citep{1999A&AS..139...25W},
\citep{2000A&A...353..958S,2002ApJ...565..539S},
\citep{2001Sci...292..698O}, the Southern Infrared Proper Motion
Survey (SIPS) \citep{SIPS1,SIPS2}, the IPHAS-POSS-I proper motion
survey of the Galactic plane \citep{2009MNRAS.397.1685D} and Lepine's
SUPERBLINK survey \citep{2005AJ....130.1247L,2008AJ....135.2177L}.

The Research Consortium On Nearby Stars (RECONS)
group\footnote{\it www.recons.org} has also been systematically canvassing
the southern sky for new proper motion systems as part of their effort
to understand the stellar population of the solar neighborhood.  To
date, these discoveries have been reported in five of the {\it The
  Solar Neighborhood} (TSN) series of papers
\citep{2004AJ....128..437H,2004AJ....128.2460H,2005AJ....129..413S,
  2005AJ....130.1658S,2007AJ....133.2898F}.  The new systems are given
SCR (SuperCOSMOS-RECONS) names because they have been discovered using
the SuperCOSMOS Sky Survey (SSS) data \citep{superc1}.  The RECONS
group continues to operate a trigonometric parallax program at the
CTIO 0.9m telescope to confirm stars within 25 pc, with a focus on
stars within 10 pc.  In Table~\ref{diststat}, we summarize the number
of new stellar systems reported to be within 25 pc by RECONS and
others via proper motion surveys that have distance estimates derived
from photographic relations.

In this investigation we focus on stars in the newly released third
U.S. Naval Observatory (USNO) CCD Astrograph Catalog (UCAC3)
\citep{u3r} found between declinations $-$90$\degr$ and $-$47$\degr$
that have 0$\farcs$40 yr$^{-1}$ $>$ $\mu$ $\ge$ 0$\farcs$18 yr$^{-1}$,
where $\mu$ is the proper motion.  The search region and proper motion
range match that in {\it The Solar Neighborhood} XVIII (hereafter,
TSN18) \citep{2007AJ....133.2898F}, in which the lower proper motion
cutoff was chosen to match that of the NLTT catalog.  In TSN18 we
presented 1606 new SCR systems, including 54 candidate common proper
motion multiples.  By utilizing the UCAC3 catalog with proper motions
determined without the sole use of photographic plates, we can probe
for new proper motion stars and companions that have been overlooked
during previous searches.  The new objects reported here have been
dubbed UPM, for this new USNO Proper Motion search.

\section {Method}
\subsection {UCAC3}

The USNO CCD Astrograph Catalog (UCAC) project has been producing
astrometric catalogs since October 2000, with the first release
(UCAC1) \citep{2000AJ....120.2131Z} covering only 80\% of the southern
sky.  The second catalog in this series (UCAC2)
\citep{2004AJ....127.3043Z} was released in July 2003 with about the
same level of completeness as UCAC1, but with early epoch plates
paired with the Astrograph CCD data for improved proper motions.  The
UCAC3, released in August 2009, is the first from the series to have
all sky coverage, and contains just over 100 million entries with a
limiting magnitude of $\sim$16 in the UCAC bandpass (579-642 nm).
UCAC3 also includes double star fitting, and has a slightly deeper
limiting magnitude than UCAC2 due to a complete re-reduction of the
pixel data \citep{u3x}.  A detailed introduction to the UCAC3 can be
found in the release paper \citep{u3r} and the README file of the data
distribution.

The UCAC3 has been used in the present survey to probe for proper
motion stars that have been overlooked during previous SCR and other
searches.  The Two-Micron All Sky Survey (2MASS) was used to probe for
and reduce systematic errors in UCAC CCD observations, giving a
greater number of reference stars to stack up residuals as a function
of many parameters, such as observing site and exposure time.  A
detailed description of the astrometric reductions of UCAC3 can be
found in \citep{u3a}.

\subsection {PROPER MOTIONS}

Out of the roughly 100 million stars in the UCAC3 catalog, about 95
million have calculated absolute proper motions.  Most proper motions
are derived using the Astrograph CCD data combined with various
earlier epoch catalogs in much the same manner as UCAC2. All input
catalogs are reduced to the International Celestial Reference Frame
(ICRF) by utilizing Hipparcos data or a similar, denser catalog, such
as Tycho-2.  For each position, standard errors are estimated.  These
errors are then used as weights to compute a UCAC3 mean position and
proper motion, utilizing a weighted, least-squares adjustment
procedure.  For bright stars (R$\sim$8--12), UCAC Astrograph CCD data
were combined with ground-based photographic and transit circle
catalogs, including all catalogs used for the Tycho-2 project
\citep{2000A&A...355L..27H}, and $\sim$1.2 million positions from
about 1950 AGK2 plates derived using the StarScan machine
\citep{starscan}.  For fainter stars (R$\sim$12.5--16.5), UCAC
Astrograph CCD data were combined with scans from the StarScan machine
of roughly 3200 plates from the Hamburg Zone Astrograph, the USNO
Black Birch Astrograph, and the Lick Astrograph, as well as a complete
new reduction of the Yale Southern Proper Motion (SPM4) survey
\citep{spm}, and data from the SuperCOSMOS project \citep{superc1}.
The SuperCOSMOS data were used in place of the Northern Proper Motions
(NPM) \citep{npm}, in preparation, which was not complete when the
UCAC3 was generated, but which will be included in the anticipated
UCAC4 final release.  An estimated error floor has been added to all
catalogs used for the proper motion calculation.  The largest root
mean square (RMS) error contribution added was 100 mas for the
SuperCOSMOS data due to zonal (plate pattern) systematic errors in the
range of 50 to 200 mas, when compared to 2MASS data.  For a detailed
description of the derived UCAC3 proper motions see \citep{u3r}.

An effort was made to tag previously known High Proper Motion (HPM)
stars in the UCAC3 catalog using the VizieR on-line data tool, along
with published literature.  The list includes roughly 51000 known
proper motion stars covering the entire sky with $\mu$ $>$
$\sim$0$\farcs$18 yr$^{-1}$.  In the North we used the LSPM-North
catalog \citep{2005AJ....130.1247L} containing 61977 new and
previously found stars having proper motions greater than 0$\farcs$15
yr$^{-1}$.  In the South we used many smaller surveys along with the
Revised NLTT Catalog \citep{2003ApJ...582.1011S}, which produced 17730
stars with proper motions greater than 0$\farcs$15 yr$^{-1}$.  For a
full list of catalogs used see the UCAC3 README file.  This list is
not comprehensive due to not having a complete list of proper motion
surveys and the dificulty in matching some catalogs like the NLTT
which do not have reliable positions. The proper motion values given
in UCAC3 for these previously known stars come from the individual
catalogs themselves and are not derived in the same manner as the
UCAC3 proper motions mentioned above.  These previously known proper
motion stars are flagged in the UCAC3 data with a Mean Position (MPOS)
running star number greater than 140 million.

The errors in proper motions reported in the UCAC3 release for stars
brighter than mag 12 are only $\sim$1--3 mas/yr because of the large
epoch spread, oftentimes as long as 100 years.  For fainter stars
found in SPM4, the errors are $\sim$2--3 mas/yr, while proper motions
incorporating SuperCOSMOS data result in errors of $\sim$6--8 mas/yr.

\subsection {SEARCH CRITERIA}

The initial sample of 177231 proper motion candidates for this search
included all UCAC3 stars in the southern sky between declinations
$-$90$\degr$ and $-$47$\degr$ with 0$\farcs$40 yr$^{-1}$ $>$ $\mu$
$\ge$ 0$\farcs$18 yr$^{-1}$.  Winnowing of the sample was accomplished
by examining previously known proper motion stars meeting the survey
criteria to find a combination of UCAC3 flags with values indicative
of real proper motion objects.  To verify the set of flags adopted for
final target selection, visual inspections were done of targets in
selected sky regions to confirm true proper motion.  In addition to
meeting the declination and proper motion survey limits, all stars (1)
must be in the 2MASS catalog with an e2mpho (2MASS photometry error)
less than or equal to 0.05 magnitudes in all three bands, (2) have a
UCAC fit model magnitude between 7 and 17 mag, (3) have a double star
flag (dsf) equal to 0, 1, 5 or 6, meaning a single star or fitted
double, (4) have an object flag (objt) between $-$2 and 2 to exclude
positions that used all overexposed images in the fit, (5) have an
MPOS number less than 140 million to exclude already known high proper
motion stars, and (6) have a LEDA galaxy flag of zero, meaning that
the source is not in the LEDA galaxy catalog.  After implementing
these cuts, 9248 candidates remained.

These candidates were then cross-checked via VizieR, the published
literature, and SIMBAD to determine if they were previously known.
VizieR was used to cross-check various proper motion catalogs, such as
NLTT, Hipparcos, and Tycho-2.  If a survey was not available on
VizieR, data were obtained from the published literature (as in the
case for stars found in TSN18, which has only recently been added to
the VizieR database.)  A final search was done by checking the
remaining candidates against the SIMBAD database.  Cross-checks of the
various compendia were performed using a 90$\arcsec$ search radius,
with one exception (the NLTT catalog, see below).  If a candidate was
matched to a known proper motion star having roughly the same proper
motion and magnitude, then it was labeled as previously known, and is
not included in the sample reported here.

A larger search radius of 180$\arcsec$ was used when comparing UPM
candidates to the NLTT catalog, which is known to have inaccurate
positions.  As shown in the Figure 2 histogram of
\citep{2002ApJS..141..187B}, the distances between their measured
positions and Luyten's listed positions of LHS stars can be quite
large.  The number of objects with a given position offset levels off
around 90$\arcsec$, beyond which fewer than 10 objects per 1$\arcsec$
bin are found.  Thus, UCAC3 proper motion candidates with positions
differing from Luyten's by less than 90$\arcsec$ are considered known,
those differing by 90--180$\arcsec$ are considered new discoveries but
are noted as possible NLTT stars in the tables, and those differing by
more than 180$\arcsec$ are considered new discoveries.  It is not a
goal of this paper to revise the NLTT catalog and assign proper
identifications and accurate positions to NLTT entries; rather, the
goal is to identify new high proper motion stars.

The various cross-checks for previously known stars reduced the number
of new candidates to a list of 4425.  Each was visually inspected to
verify proper motion by blinking the $B_J$ and $R_{59F}$ SuperCOSMOS
digitized plate images.  Objects without verifiable proper motions
were then discarded, leaving 474 new proper motion discoveries.  Of
these new discoveries 32 were found to be part of a Common Proper
Motion (CPM) system, including seven new discoveries having CPM to
previously known primaries, leaving a total of 442 new systems.

A lower successful hit rate (5297 real proper motion objects / 9248
total ``good'' candidates extracted) of 57.3\% was found for this
search than the 78.1\% successful hit rate obtained in TSN18. As in
TSN18, the hit rate takes into account new, known, and phantom proper
motion objects (phantoms are identified as moving objects but are
not).  The lower hit rate in the present effort is the result of at
least three factors.  First, some real objects were discarded early in
the search due to the rigorous sifting mentioned above to obtain a
more manageable sample for investigating, i.e. some of the selection
criteria, particularly involving 2mass, were ``too tight.''  Second,
many phantom proper motion objects in the UCAC3 made the sample cuts
because of incorrect matches between catalogs during the proper motion
calculation.  This is particularly common in the fainter stars for
which the proper motion calculations rely on only two catalog
positions.  Third, other misidentifications arise from blended images,
where two single star detections in the UCAC3 can be matched up to a
single image in an earlier epoch catalog.

\section {RESULTS}

The 442 new UCAC3 proper motion systems are listed in
Table~\ref{U3-discoveries}.  In Table~\ref{U3-distance} we highlight
the 15 systems for which we estimate distances to be less than 25 pc.
In both Tables we list names, coordinates, proper motions, 1$\sigma$
errors in the proper motions, plate magnitudes from SuperCOSMOS,
near-IR photometry from 2MASS, the computed $R_{59F}-J$ color,
distance estimate, and notes.

\subsection{Positions and Proper Motions}

All positions on the ICRF system, proper motions, and errors are taken
directly from UCAC3.  For a few stars found visually, e.g. companions,
no data could be obtained from UCAC3, so information was obtained from
other sources (additional objects, see below).  The average positional
errors reported in the UCAC3 catalog for this sample are 52 mas in RA
and 53 mas in Dec.  The average proper motion errors are 8.5 mas/yr in
$\mu_{\alpha}\cos\delta$ and 8.4 mas/yr in $\mu_{\delta}$.

\subsection{Photometry}

In Tables~\ref{U3-discoveries} and \ref{U3-distance}, we give
photographic magnitudes from SuperCOSMOS for three plate emulsions:
$B_J$, $R_{59F}$, and $I_{IVN}$.  For sources fainter than mag
$\sim$15, plate errors are typically less than 0.3 mag, but errors
increase for brighter sources.  Plate color errors are smaller, at
roughly 0.07 mag \citep{2001A&A....326..1295H}.  2MASS $JHK_s$
infrared photometry is given, with errors typically 0.05 mag or less
due to the search criteria and because the stars are usually brighter
than 16 in UCAC3 and are red, making them relatively bright, mag
$\sim$ 10--14, in 2MASS.  Companions found during visual inspection
may be fainter and have consequently larger photometric errors.  The
optical (SuperCOSMOS) and infrared (2MASS) datasets are combined in a
computed $R_{59F}-J$ color to provide an indicator of the star's
color.  In some cases, SuperCOSMOS magnitudes may not be given due to
blending, no source detection or other problems where no magnitude is
reported in the SuperCOSMOS data.  2MASS magnitudes are given for all
but five objects that are not present in the 2MASS catalog.

\subsection{Distances}

Distance estimates are computed using 11 colors generated from the
six-band photometry using the relations given in
\citep{2004AJ....128..437H}.  This method assumes that all objects are
main sequence stars.  The accuracy reported for this technique is
roughly 26\%, which is determined from the mean of the absolute values
of the differences between distances for stars with trigonometric
parallaxes and distances estimated via the relations.  No distance
estimate is given for stars that are too blue for the relations.  For
objects with incomplete photometry, the distance estimate will be less
reliable.  While only one relation is needed to produce a distance
estimate, six are needed to be considered ``reliable.'' This is half
of the 11 total posibilities.  Stars having fewer than six relations
have been identified in the notes.  For any star expected to have an
erroneous distance (white dwarf, evolved star, subdwarf), the distance
is given in brackets.

\subsection{Additional Objects}

During the visual inspection of the candidates, 27 additional proper
motion objects were found, listed in Table 4.  These objects generally
are CPM companion candidates that either have a fainter limiting
magnitude than implemented for this search, were eliminated from the
candidate list by the search criteria, or have a UCAC3 proper motion
less than 0$\farcs$18 yr$^{-1}$.  Proper motions from the UCAC3 data
less than the 0$\farcs$18 yr$^{-1}$ cutoff of this paper are
considered suspect from a visual inspection that compared the proper
motion of the companion candidate.  All objects detected during the
visual inspection were investigated using SIMBAD and VizieR for
previous identifications.  If none were found, their proper motions
were obtained from UCAC3, SPM4, or SuperCOSMOS, in that order of
priority.  Magnitudes were then obtained from SuperCOSMOS and 2MASS to
compute distance estimates.  Only four of the 27 objects found
visually did not have a proper motion reported in any catalog.  For
stars that were not found in the UCAC3 data, positions were computed
using the epoch, coordinates, and proper motion obtained from the
corresponding catalog.

\section {ANALYSIS}

\subsection {Color-Magnitude Diagram}

\begin{figure}
  \epsscale{1.00}
  \includegraphics[angle=-90,scale=0.40]{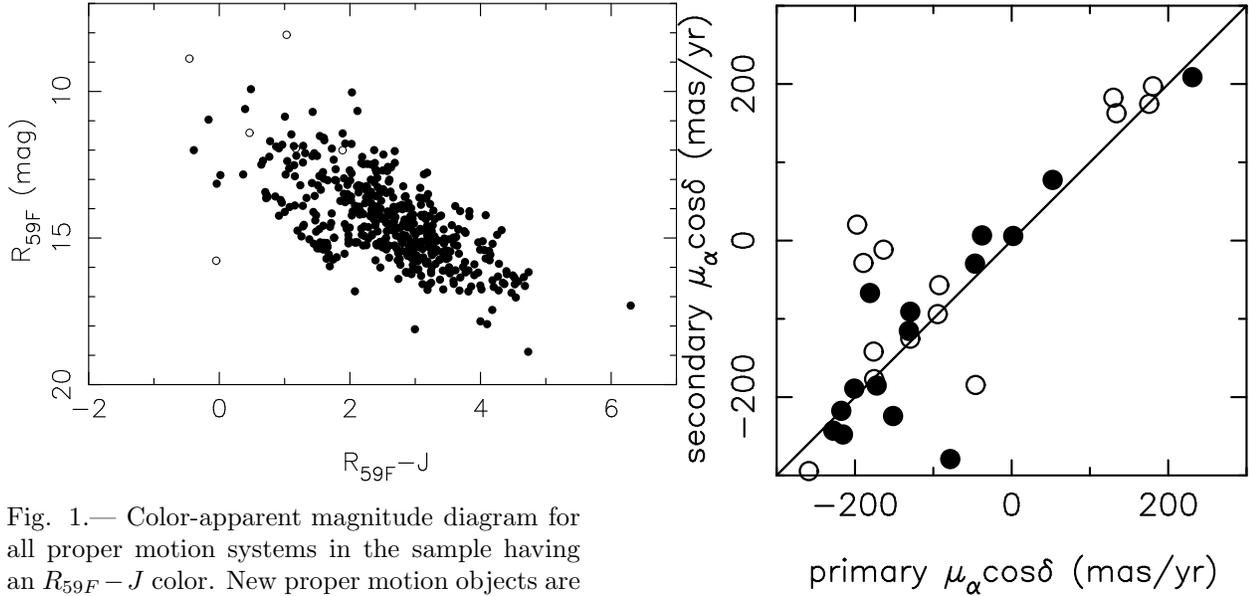}
  \caption{Color-apparent magnitude diagram for all proper motion
  systems in the sample having an $R_{59F} - J$ color.  New proper
  motion objects are represented by closed circles while known objects
  (CPM companions to new objects) are represented with open circles.
  Data points below $R_{59F} =$ 17 are CPM candidates noticed during the
  visual inspection.}\label{color}
  \end{figure}

In Figure~\ref{color} we show a color-magnitude diagram of the 465
proper motion objects reported in this sample having a $R_{59F} - J$
color.  New proper motion objects are represented by closed circles
while known objects (companions to new objects) are represented by
open circles.  Data points that fall below $R_{59F} \sim 17$ are CPM
companion candidates noticed during visual inspection.  The brightest
new object, UPM 1542-5041, has $R_{59F}$ = 9.924 and is estimated to be
at a distance of 33.1 pc.  The reddest object found in this search is
UPM 1703-4934B with $R_{59F} - J$ = 6.30, $R_{59F}$ = 17.31, and
estimated distance of 40.7 pc.

The subdwarf population is not as well defined in this paper as in
TSN18 because there are far fewer new objects.  Nonetheless, a
separation can be seen below the concentration of main sequence stars.
Finally, a single known white dwarf, WD 0607-530B, can be seen in the
lower left of Figure~\ref{color}.

\subsection {Reduced Proper Motion Diagram}

\begin{figure}
  \epsscale{1.00}
  \includegraphics[angle=-90,scale=0.40]{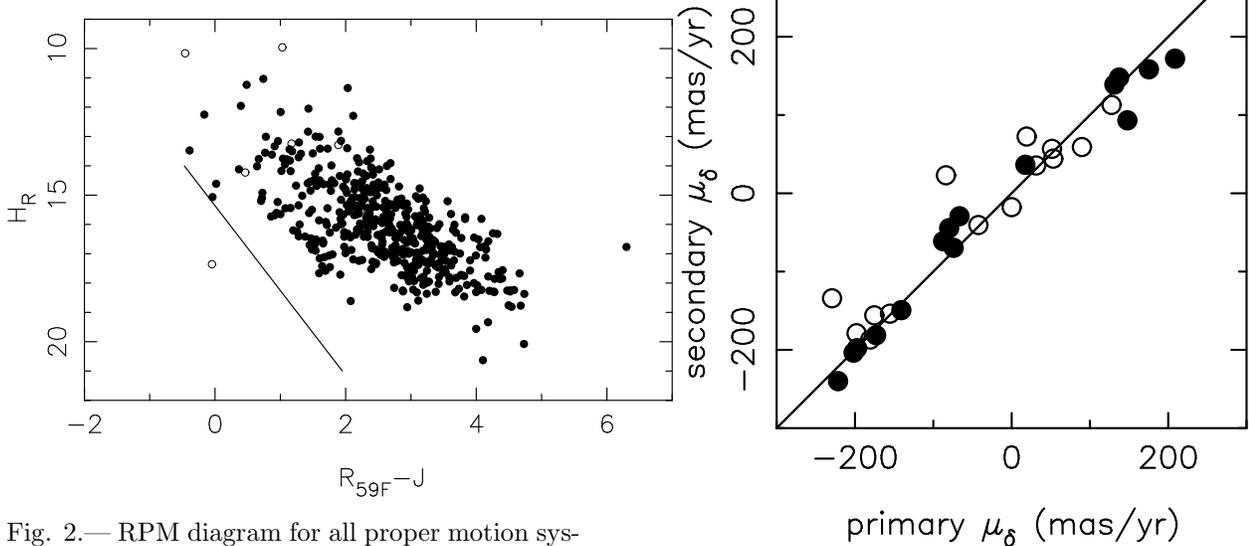}
  \caption{RPM diagram for all proper motion systems in this sample
  having an $R_{59F} - J$ color.  New proper motion objects are
  represented by closed circles while known objects (CPM companions to
  new objects) are represented with open circles. The empirical line
  separates the white dwarf candidate from the subdwarf candidates,
  which lie above the white dwarf stars and just below the
  concentration of main sequence stars.}\label{rpm}
  \end{figure}

We show in Figure~\ref{rpm} the Reduced Proper Motion (RPM) diagram
for all 465 objects in this sample having a $R_{59F} - J$ color.  New
proper motion objects are represented by closed circles while known
objects (companions to new objects) are represented by open circles.
A reduced proper motion diagram takes advantage of the assumption that
objects with larger distances tend to have smaller proper motions.
While this is not always valid it can be used as a good method to
separate white dwarfs and subdwarfs from main-sequence stars.  We
determine H$_{R}$ as in TSN18 using a modified distance modulus
equation, in which $\mu$ is substituted for distance.

\begin{displaymath}
H_R = R_{59F} + 5 + 5\log\mu.
\end{displaymath}

The dashed line in Figure~\ref{rpm} is the same empirical line used in
TSN18 to separate white dwarfs from subdwarfs.  This separation line
has been shown from past SCR searches to be reliable in identifying
white dwarf candidates.  From the present survey, only one known white
dwarf WD 0607-530B, a CPM companion candidate to UPM 0608-5301A, is
seen clearly below the subdwarf region.

Subdwarf candidates have been selected using the same method as in
TSN18 --- stars with $R_{59F} - J >$ 1.0 and within 4.0 mag in $H_{R}$
of the the empirical line separating the white dwarfs are considered
subdwarfs.  From this survey there are 31 subdwarf candidates, all
with distance estimates greater than 147 pc.  Large distance estimates
can be used to identify both subdwarf and white dwarf candidates,
which are subluminous compared to main sequence stars and yield large
distance estimates because they are intrinsically fainter than the
main-sequence stars used to generate the photometric distance
relations.  The presumably erroneous distances for these stars are
given in brackets in Tables~\ref{U3-discoveries} and
\ref{U3-distance}.  Follow up spectroscopic observations will be
needed to confirm all subdwarf candidates.

\subsection {New Common Proper Motion Systems}

In this search we found 32 common proper motion systems (31 binaries
and one triple), including 25 entirely new systems and seven hybrid
systems containing both new and known objects.  The triple system is a
previously known system discovered as part of the automated search to
have a newly discovered third component.  The data for these systems
are given in Table~\ref{U3CPM}, where we list the primaries and
companions, their proper motions, and the companions' separations and
position angles relative to the primaries (defined to be the brightest
star in each system).  The distance estimates were used in conjunction
with the proper motions and visual inspections to determine whether or
not a pair of stars is physically associated.  Because these objects
were found during visual inspections, the proper motion and/or
SuperCOSMOS magnitudes may be missing or suspect; in such cases,
identifications as CPM systems are more tentative and identified in
the notes.

\begin{figure}
  \epsscale{1.00}  
  \plotone{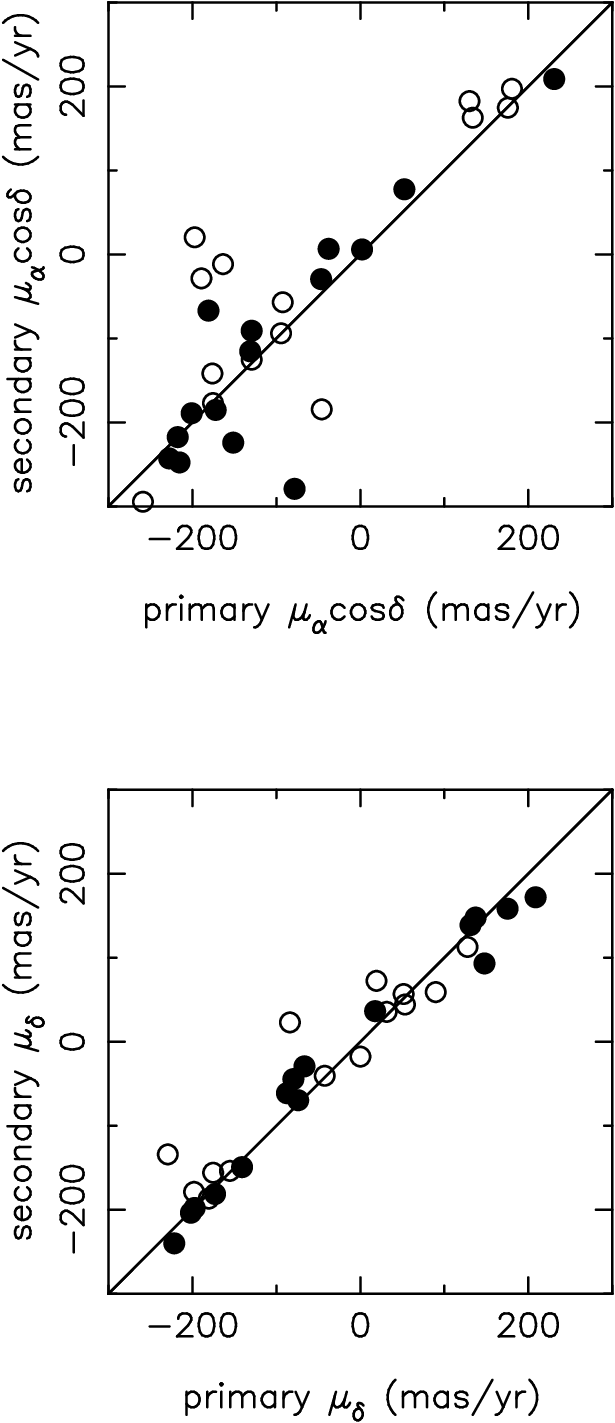} 
  \caption{Comparison of proper motions per coordinate,
    $\mu_{\alpha}\cos\delta$ (top) and $\mu_{\delta}$ (bottom), for
    components in CPM systems. Proper motions from the UCAC3 catalog are
    represented by closed circles while proper motions manually obtained
    through other means are denoted by open circles.  The solid line
    indicates perfect agreement.}\label{cpm1}
  \end{figure}

In Figure~\ref{cpm1} we compare the proper motions per coordinate for
the 29 CPM systems for which both components have proper motions.  CPM
candidates that have proper motions from UCAC3 are represented by
closed circles while those with proper motions from other sources are
represented by open circles.  Proper motions for the latter candidates
were extracted manually from either SPM4 or SuperCOSMOS.

\subsection{Notes on Specific Stars}

{\bf UPM 0608-5301A} is an M dwarf at an estimated distance of 37.1 pc
with a known white dwarf as a possible companion.  The B component
(known white dwarf) is at a separation of 21.5$\arcsec$ at position
angle 120.7$^{\circ}$ from the primary.  We estimate a distance of
34.2 pc for the white dwarf with an error of 20\% using the relation
of \citep{2001Sci...292..698O}. See Table~\ref{U3CPM} for more
details.

{\bf UPM 0835-6018C} is in a possible triple system with NLTT 19906
and NLTT 19907.  The A and B components are separated by 5.1$\arcsec$.
The C component has a separation of 113.0$\arcsec$ at a position angle
of 49.3$^{\circ}$ from the primary. See Table~\ref{U3CPM} for more
details.

{\bf UPM 1230-5736AB} The A component is estimated to be at 22.2 pc,
and has $R_{59F} =$ 12.04 and proper motion per coordinate
($\mu_{\alpha}\cos\delta$,$\mu_{\delta}$) = (-227.6,-66.5) mas/yr.
NLTT 30961 is found 1.61$\arcmin$ away, and NLTT lists a red
photographic magnitude of 13.1 and proper motion per coordinate
($\mu_{\alpha}\cos\delta$,$\mu_{\delta}$) = (-216.7, -38.2) mas/yr.

The B component is estimated to be at 19.8 pc, and has $R_{59F} =$
12.83 and proper motion per coordinate
($\mu_{\alpha}\cos\delta$,$\mu_{\delta}$) = (-243.0, -29.3) mas/yr.
NLTT 30938 is found 1.68$\arcmin$ away, and NLTT lists a red
photographic magnitude of 12.6 and proper motion identical to NLTT
30961.  Thus, this nearby UPM double is likely the same as the NLTT
double, but the relatively large offset from the NLTT makes the
identification ambiguous.

{\bf UPM 1542-5041} is the brightest new HPM discovery from this
effort.  It has a distance estimate of 33.1 pc, $R_{59F} =$ 9.92 and
proper motion per coordinate ($\mu_{\alpha}\cos\delta$,$\mu_{\delta}$)
= (182.4, -16.3) mas/yr.  NLTT 40903 is found 2.46$\arcmin$ away, and
NLTT lists a red photographic magnitude of 12.8 and proper motion per
coordinate ($\mu_{\alpha}\cos\delta$,$\mu_{\delta}$) = (-244.2,-190.8)
mas/yr.  The discordant magnitudes and proper motions indicate that
UPM 1542-5041 is not NLTT 40903.

{\bf UPM 1710-5300} has an estimated distance of only 13.5 pc, making
it the nearest candidate in the sample.

\subsection {COMPARISON TO PREVIOUS PROPER-MOTION SURVEYS}

Most previously known HPM stars have been tagged in UCAC3 and their
listed proper motions in UCAC3 were taken from their respective
catalogs.  Because no UCAC3 proper motions were determined,
comparisons to other catalogs/surveys are therefore difficult.
Nonetheless, within the sky coverage and proper motion regime of this
paper, 66 stars have been found in both the Hipparcos and Tycho-2
catalogs that are not tagged as HPM stars in the UCAC3 catalog.  This
constitutes a small but ample number of stars that can be used to
compare the bright end of UCAC3 proper motions to those in the
Hipparcos and Tycho-2 catalogs.  In Figure~\ref{pm1}, we show the
comparison between UCAC3 proper motions in RA and DEC to the Hipparcos
(top) and Tycho-2 (middle) catalogs.  For comparison we also plot the
proper motion differences between Hipparcos and Tycho-2 in the bottom
panel of Figure~\ref{pm1}.  This plot implies that for both the
Hipparcos and Tycho-2 catalogs the UCAC3 proper motions show only
small differences per coordinate at all declinations in the present
search, with no significant systematics.  The RMS differences of UCAC3
proper motions per coordinate ($\Delta\mu_{\alpha}\cos\delta$,
$\Delta\mu_{\delta}$) when compared to Hipparcos are 7.5 and 6.6
mas/yr, respectively.  When compared to Tycho-2 proper motion
coordinates we find 5.5 and 5.9 mas/yr, respectively.  A slightly
lower RMS difference of 3.8 mas/yr in both coordinates is seen when
comparing the Hipparcos and Tycho-2 proper motions for these stars.

To investigate fainter stars in UCAC3, we compare UCAC3 proper motions
to SPM4 and SuperCOSMOS results.  We compare proper motions using the
SuperCOSMOS proper motions to bring the positions to the UCAC3 epoch
with a 1.5 arcsec match radius.  A 1.5 arcsec radius was also used to
match UCAC3 to SPM4 with no need to correct for proper motions because
both catalogs are on the same system and set at the same epoch.

A total of 137 objects meeting the proper motion and declination
limits of this paper were found in all three catalogs.  In
Figure~\ref{pm2}, we compare UCAC3 proper motions in the same manner
as above with the SPM4 (top) and SuperCOSMOS (middle) catalogs.  Again
for comparison, we also include a plot showing the differences between
SPM4 and SuperCOSMOS in the bottom panel of Figure~\ref{pm2}.  The RMS
differences between UCAC3 and SPM4 per coordinate
($\Delta\mu_{\alpha}\cos\delta$, $\Delta\mu_{\delta}$) are 6.6 and 4.1
mas/yr respectively.  Much higher RMS differences of 19.3 and 18.9
mas/yr are seen when comparing UCAC3 to the SuperCOSMOS proper motions
per coordinate ($\Delta\mu_{\alpha}\cos\delta$, $\Delta\mu_{\delta}$).
This comparison also indicates that proper motions in RA are
systematically shifted in the SuperCOSMOS data, but are consistent in
DEC.  This high RMS including the systematic shift in RA is also seen
when comparing the SPM4 to SuperCOSMOS proper motions per coordinate,
yielding RMS differences of 19.9 and 17.5 mas/yr in
$\Delta\mu_{\alpha}\cos\delta$ and $\Delta\mu_{\delta}$, respectively.
The higher RMS differences for the SuperCOSMOS proper motions are
in agreement with the findings of TSN18 where SCR proper motions were
found to have an average deviation of 23 mas/yr total proper motion
when compared to the NLTT and Hipparcos proper motions.

In TSN18 a total of 1662 objects were reported, of which 1615 match
the proper motion and declination limits of this paper.  During this
UCAC3 search, 1298 of the 1615 objects reported in TSN18 were
recovered, or a 80.4\% succesful recovery rate.  Objects missed in
this UCAC3 survey are primarily those at the faint end, as the TSN18
survey reached to $R_{59F} =$ 16.5.

The Hipparcos catalog contains 118218 total objects of which 722 meet
the proper motion and declination limits of this paper.  Tycho-2
contains 2539913 total objects in the main catalog with 1273 of those
objects matching the proper motion and declination limits of this
paper.  We recover 646 Hipparcos stars and 973 Tycho-2 stars using the
search criteria of this paper, yielding recovery rates of 89.5\% and
76.4\% respectively.  Objects missed in this UCAC3 survey is primarily
due to UCAC3 lacking a source detection for $\sim$15\% of the Tycho-2
objects.  The relatively high recovery rates of UCAC3 when compared to
these three efforts implies UCAC3 can be used as a reliable source to
search for new proper motion stars with $\mu$ = 0.18--0.40 arcsec
yr$^{-1}$ for other portions of the sky.

\section {DISCUSSION}

We have found 442 new proper motion systems including 474 objects with
0$\farcs$40 yr$^{-1}$ $>$ $\mu$ $\ge$ 0$\farcs$18 yr$^{-1}$ between
declinations $-$90$\degr$ and $-$47$\degr$.  In Figure~\ref{sky}, we
show the sky distribution for the entire sample reported in this
paper.  The 474 new discoveries represent a 25.3\% increase in new
systems for the same region of the sky covered by previous (SCR)
searches that used Schmidt plates as the primary source of discovery.
While many of these new UPM discoveries are found along the Galactic
plane, a region avoided by the SCR survey, additional new systems were
found far from the plane.  Areas in Figure~\ref{sky} with a lower
density of new discoveries have been heavily searched by previous
proper motion surveys as seen when comparing to a similar sky plot
presented in TSN18.

\begin{figure}
  \epsscale{1.00}  
  \plotone{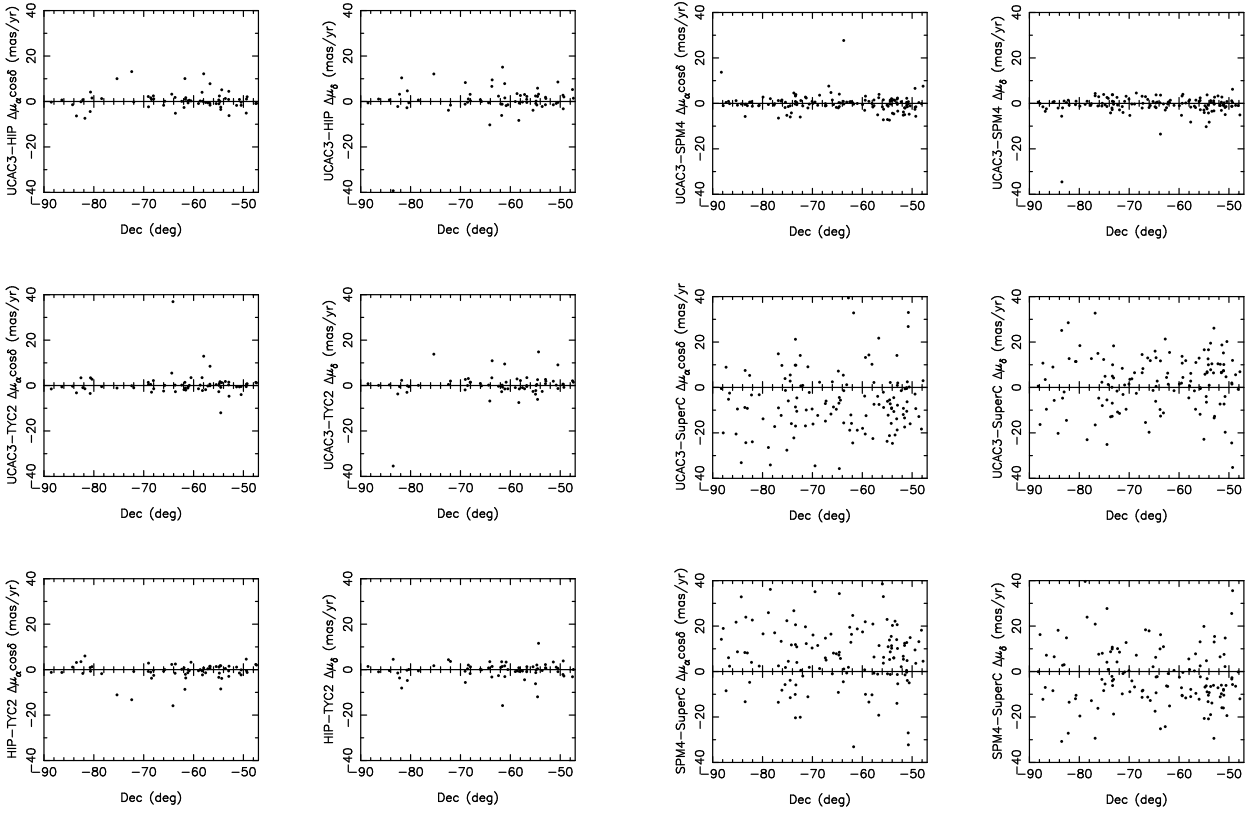} 
  \caption{Comparison of UCAC3, Hipparcos and Tycho-2 proper motions per
    coordinate, $\Delta\mu_{\alpha}\cos\delta$ (left column) and
    $\Delta\mu_{\delta}$ (right column). }\label{pm1}
  \end{figure}

\begin{figure}
  \epsscale{1.00}  
  \plotone{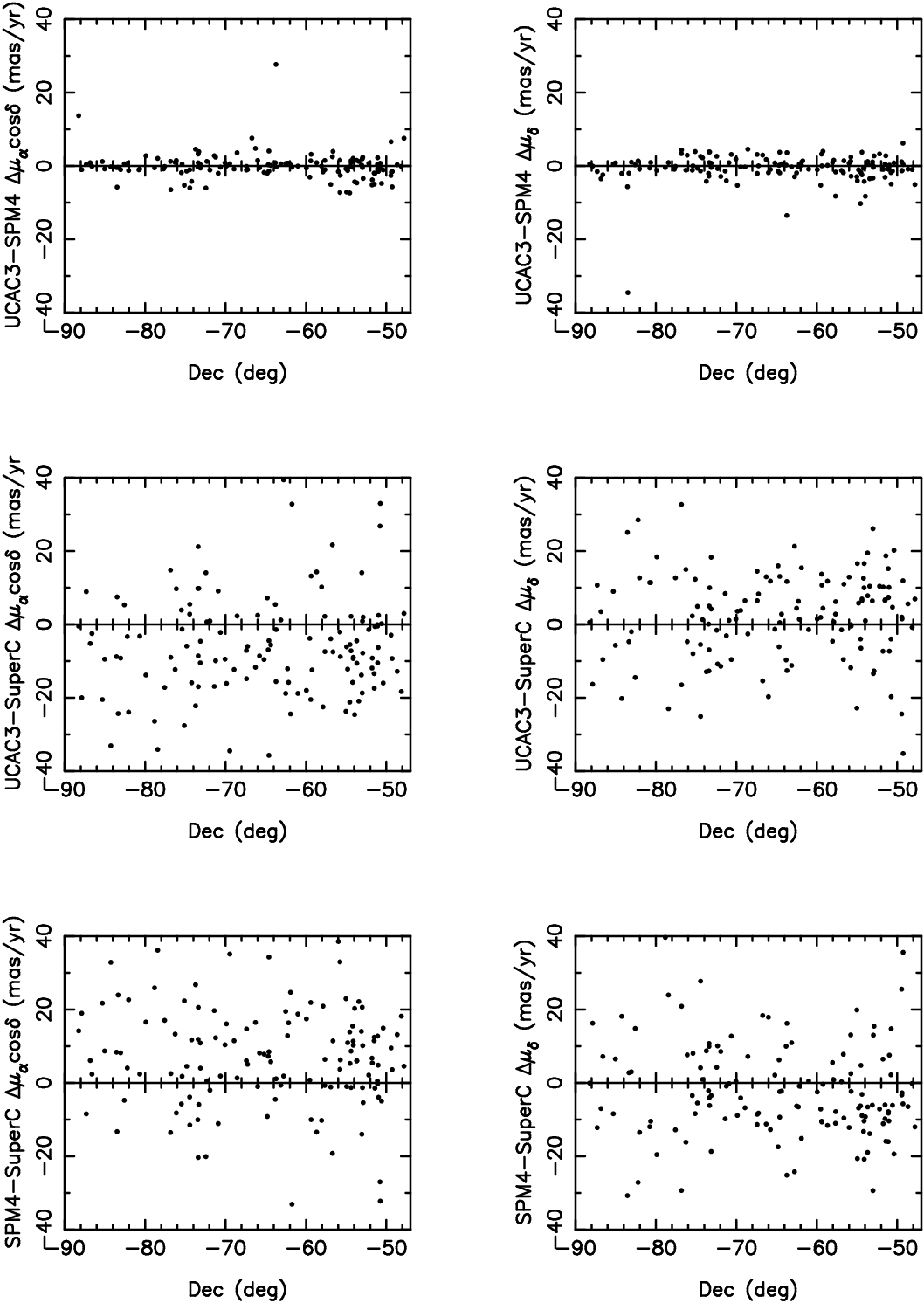} 
  \caption{Comparison of UCAC3, SuperCOSMOS and SPM4 proper motions per
    coordinate, $\Delta\mu_{\alpha}\cos\delta$ (left column) and
    $\Delta\mu_{\delta}$ (right column). }\label{pm2}
  \end{figure}

\begin{figure}
  \epsscale{1.00}
  \includegraphics[angle=-90,scale=0.40]{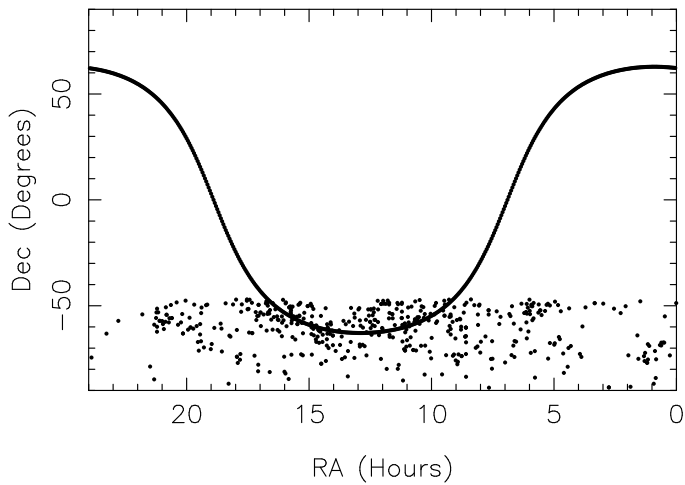}
  \caption{Sky distribution of all UCAC3 proper motion objects
  reported in this sample, i.e.~those between declinations
  $-$90$\degr$ and $-$47$\degr$ having 0$\farcs$40 yr$^{-1}$ $>$
  $\mu$ $\ge$ 0$\farcs$18 yr$^{-1}$.  The curve represents the
  Galactic plane. }\label{sky}
  \end{figure}

As shown in Figure~\ref{pm1} and \ref{pm2}, the proper motions
obtained from UCAC3 compare well to the Hipparcos, Tycho-2, and SPM4
catalogs.  However, we find that the SuperCOSMOS proper motions have a
significantly higher scatter when compared to these catalogs, which
confirms our similar result in TSN18.

\begin{figure}
  \epsscale{1.00}  
  \includegraphics[angle=-90,scale=0.40]{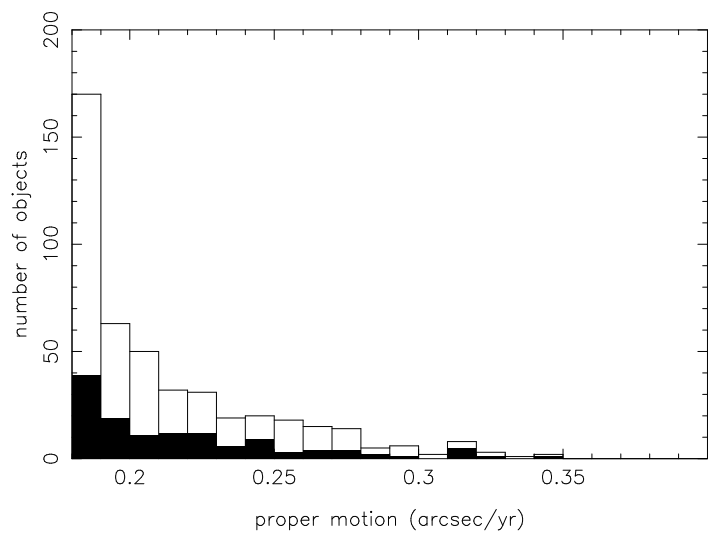}
  \caption{Histogram showing the number of proper motion objects in
  0$\farcs$01 yr$^{-1}$ bins for the entire sample (empty bars) and
  the number of those objects having distance estimates within 50 pc
  (filled bars).
  }\label{hist}
  \end{figure}

We find 25 new CPM candidate systems, as well as 31 new subdwarf
candidates that will need future spectroscopic efforts to be
confirmed.  In Figure~\ref{hist}, we show a histogram of the number of
proper motion discoveries in 0$\farcs$01 yr$^{-1}$ bins for the
present sample, highlighting the number of those having distance
estimates within 50 pc.  The increase in nearby systems at the lowest
proper motions sampled here implies that more nearby stars are likely
to be found at even slower proper motion regimes.

Finally, we have found 16 objects in 15 systems with distances
estimated to be within 25 pc, and an additional 109 objects in 107
systems between 25 and 50 pc.  The discoveries include UPM 1542-5041,
which at $R_{59F} =$ 9.92 is a surprisingly bright new proper motion
discovery with an estimated distance of 33.1 pc.  UPM 1710-5300, which
is our nearest new candidate with an estimated distance at 13.5 pc.
We anticipate that further exploration of the UCAC3 for new proper
motion discoveries will result in more nearby star candidates, perhaps
some even within 10 pc, where new discoveries are still being made
\citep{2006AJ....132.2360H}
 

\acknowledgments

We thank the entire UCAC team for making this proper motion survey
possible.  Special thanks go out to the RECONS team at Georgia State
University for their support, John Subasavage in particular for
assistance with the SCR searches, and Nigel Hambly for his work with
the SuperCOSMOS Sky Survey.  We would also like to thank all USNO
summer students who helped in this survey.  This work has made use of
the SIMBAD, VizieR, and Aladin databases operated at the CDS in
Strasbourg, France.  We have also made use of data from the Two-Micron
All Sky Survey, SuperCOSMOS Science Archive and the Southern Proper
Motion catalog.


\clearpage


  \clearpage
 
  \clearpage
  
 \clearpage



\begin{deluxetable}{lcl}
\tabletypesize{}

\tablecaption{New Proper Motion Systems within 25 Parsecs Based on
  Distances Using SuperCOSMOS Photographic Magnitudes\label{diststat}}

\tablewidth{0pt}
\tablehead{
\colhead{Paper}             &
\colhead{Systems}           &
\colhead{References}       \\

\colhead{}                  &
\colhead{$\le$ 25 pc}       &
\colhead{}                  }

\startdata
  TSN08      &   3  & \citep{2004AJ....128..437H} \\
  TSN10      &   4  & \citep{2004AJ....128.2460H} \\
  TSN12      &  18  & \citep{2005AJ....129..413S} \\
  TSN15      &  25  & \citep{2005AJ....130.1658S} \\
  TSN18      &  31  & \citep{2007AJ....133.2898F} \\
             &      &                             \\
  SIPS1      &   1  & \citep{SIPS1}               \\
  SIPS2      &  12  & \citep{SIPS2}               \\
             &      &                             \\
  this paper &  15  & (Finch et al. 2010)         \\
\enddata
\end{deluxetable}
\clearpage

\begin{deluxetable}{lrrrrrrrrrrrrrrl}
\tabletypesize{\tiny} 
\rotate

\tablecaption{New UCAC3 High Proper Motion Systems between
  Declinations $-$90$\degr$ and $-$47$\degr$ with 0$\farcs$40
  yr$^{-1}$ $>$ $\mu$ $\ge$ 0$\farcs$18
  yr$^{-1}$ \label{U3-discoveries}} \tablewidth{0pt}

\tablehead{
\colhead{Name}                         &
\colhead{RA}                           &
\colhead{DEC}                          &
\colhead{$\mu_{\alpha}\cos\delta$}     &
\colhead{$\mu_{\delta}$}               &
\colhead{sig$\mu_{\alpha}$}            &
\colhead{sig$\mu_{\delta}$}            &
\colhead{$B_J$}                        &
\colhead{$R_{59F}$}                    &
\colhead{$I_{IVN}$}                    &
\colhead{$J$}                          &
\colhead{$H$}                          &
\colhead{$K_s$}                        &
\colhead{$R_{59F}$ $-$ $J$}            &
\colhead{Est Dist}                     &
\colhead{Notes}                        \\
			               
\colhead{}                             &
\colhead{(mas)}                        &
\colhead{(mas)}                        &
\colhead{(mas/yr)}                     &
\colhead{(mas/yr)}                     &
\colhead{(mas/yr)}                     &
\colhead{(mas/yr)}                     &
\colhead{}                             &
\colhead{}                             &
\colhead{}                             &
\colhead{}                             &
\colhead{}                             &
\colhead{}                             &
\colhead{}                             &
\colhead{(pc)}                         &
\colhead{}                             }

\startdata
 UPM 0000-4843  &     480848 & 148604159 &   181.7 &    29.9 &    12.3 &     8.5 &  15.579 &  13.333 &  11.337 &  10.824 &  10.208 &  10.018 &   2.509 &    47.2  &                          \\  
 UPM 0016-7327  &   14557586 &  59575305 &   184.8 &   -29.6 &     1.9 &     2.1 &  16.498 &  15.410 &  13.572 &  11.351 &  10.754 &  10.611 &   4.059 &    40.0  &                          \\  
 UPM 0028-5501  &   25583309 & 125920383 &   190.0 &   -10.2 &     2.6 &     3.0 &  17.382 &  15.533 &  14.688 &  11.863 &  11.233 &  11.054 &   3.670 &    50.1  &                          \\ 
 UPM 0029-6724  &   26938223 &  81322056 &   184.7 &    43.5 &     1.8 &     1.4 &  15.228 &  13.197 &  12.132 &  11.065 &  10.412 &  10.244 &   2.132 &    60.7  &                          \\  
 UPM 0032-6719  &   29092633 &  81607463 &   183.1 &    90.8 &     8.9 &     5.5 &  17.735 &  15.518 &  14.025 &  12.776 &  12.160 &  11.956 &   2.742 &   100.6  &                          \\  
 UPM 0037-6725  &   33311860 &  81260513 &   297.2 &   -98.6 &    11.2 &    11.8 &  17.318 &  15.043 &  13.494 &  11.529 &  11.025 &  10.739 &   3.514 &    40.5  &                          \\  
 UPM 0037-6844  &   33623179 &  76513673 &   249.0 &    44.0 &     7.8 &     7.6 &  17.173 &  15.016 &  13.408 &  11.571 &  10.983 &  10.701 &   3.445 &    41.1  &                          \\  
 UPM 0040-6834  &   36433102 &  77158520 &   170.0 &    78.5 &     7.4 &     7.1 &  15.386 &  13.502 &  12.456 &  11.054 &  10.405 &  10.257 &   2.448 &    55.1  &                          \\
\enddata

\tablenotetext{a}{Subdwarf candidate picked from RPM diagram; plate distance [in bracket] is incorrect}
\tablenotetext{b}{Number of relations used for distance estimate $<$ 6: plate distance less reliable}
\tablenotetext{c}{Common proper motion companion; see Table~\ref{U3CPM}}
\tablenotetext{d}{Not detected during automated search but noticed by eye during the blinking process}
\tablenotetext{e}{Proper motions suspect}
\tablenotetext{f}{Possible NLTT star with a position difference $>$ 90$\arcsec$ when compared to UCAC3 position}
\tablenotetext{g}{Source not in 2MASS}
\tablenotetext{h}{SuperCOSMOS plate magnitudes suspect}
\end{deluxetable}


\begin{deluxetable}{lrrrrrrrrrrrrrrl}
\tabletypesize{\tiny} 
\rotate 

\tablecaption{New UCAC3 High Proper Motion Systems estimated to be
  within 25 pc between Declinations $-$90$\degr$ and $-$47$\degr$ with
  0$\farcs$40 yr$^{-1}$ $>$ $\mu$ $\ge$ 0$\farcs$18
  yr$^{-1}$ \label{U3-distance}} \tablewidth{0pt}

\tablehead{
\colhead{Name}                          &
\colhead{RA}                            &
\colhead{DEC}                           &
\colhead{$\mu_{\alpha}\cos\delta$}      &
\colhead{$\mu_{\delta}$}                &
\colhead{sig $\mu_{\alpha}$}            &
\colhead{sig $\mu_{\delta}$}            &
\colhead{$B_J$}                         &
\colhead{$R_{59F}$}                     &
\colhead{$I_{IVN}$}                     &
\colhead{$J$}                           &
\colhead{$H$}                           &
\colhead{$K_s$}                         &
\colhead{$R_{59F}$ $-$ $J$}             &
\colhead{Est Dist}                      &
\colhead{Notes}                         \\
			                
\colhead{}                              &
\colhead{(mas)}                         &
\colhead{(mas)}                         &
\colhead{(mas/yr)}                      &
\colhead{(mas/yr)}                      &
\colhead{(mas/yr)}                      &
\colhead{(mas/yr)}                      &
\colhead{}                              &
\colhead{}                              &
\colhead{}                              &
\colhead{}                              &
\colhead{}                              &
\colhead{}                              &
\colhead{}                              &
\colhead{(pc)}                          &
\colhead{}                              }

\startdata
 UPM 0559-5225  &  323598169 & 135295015 &   120.7 &   152.6 &     4.5 &     4.4 &  14.532 &  12.000 &  10.287 &   9.628 &   8.951 &   8.703 &   2.372 &    24.9  &                                       \\ 
 UPM 0621-6111  &  343586117 & 103726637 &    11.4 &   204.6 &     5.8 &     5.3 &  16.108 &  14.256 &  12.028 &  10.427 &   9.788 &   9.534 &   3.829 &    21.8  &                                       \\ 
 UPM 0901-6526  &  487375530 &  88399668 &   -68.4 &   196.4 &     6.7 &     6.7 &  16.269 &  14.219 &  11.471 &  10.140 &   9.589 &   9.282 &   4.079 &    18.0  &                                       \\ 
 UPM 1104-6232  &  598106935 &  98845495 &  -207.8 &   -58.5 &     6.0 &     6.2 &  16.423 &  14.086 &  11.484 &  10.256 &   9.677 &   9.357 &   3.830 &    19.4  &                                       \\ 
 UPM 1105-5825A &  598956034 & 113681441 &  -176.1 &    51.5 &     4.2 &     4.2 &  16.916 & \nodata &  13.713 &  10.298 &   9.711 &   9.496 & \nodata &    15.2  & \tablenotemark{a},\tablenotemark{b}   \\ 
 UPM 1142-6440  &  632248846 &  91165839 &  -183.4 &  -205.1 &    32.2 &    16.8 &  16.067 &  14.081 &  11.418 &  10.393 &   9.846 &   9.481 &   3.688 &    23.7  &                                       \\ 
 UPM 1230-5736B &  675249073 & 116560461 &  -243.0 &   -29.3 &     6.9 &     6.5 &  14.928 &  12.828 &  10.998 &   9.694 &   9.033 &   8.785 &   3.134 &    19.8  & \tablenotemark{a},\tablenotemark{c}   \\ 
 UPM 1230-5736A &  675783245 & 116623954 &  -227.6 &   -66.5 &     6.8 &     6.5 &  13.794 &  12.035 &  10.550 &   9.348 &   8.712 &   8.445 &   2.687 &    22.2  & \tablenotemark{a},\tablenotemark{c}   \\
\enddata

\tablenotetext{a}{Common proper motion companion; see Table~\ref{U3CPM}}
\tablenotetext{b}{SuperCOSMOS plate magnitudes suspect}
\tablenotetext{c}{Possible NLTT star with a position difference $>$ 90$\arcsec$ when compared to UCAC3 position}

\end{deluxetable}


\begin{deluxetable}{lccclcccccl}
\rotate 
\tabletypesize{\scriptsize} 
\tablecaption{Common Proper Motion Candidate
Systems\label{U3CPM}}
\tablewidth{0pt}

\tablehead{
\colhead{Primary}                        &
\colhead{$\mu_{\alpha}\cos\delta$}       &
\colhead{$\mu_{\delta}$}                 &
\colhead{Distance}                       &
\colhead{Secondary/Tertiary}             &
\colhead{$\mu_{\alpha}\cos\delta$}       &
\colhead{$\mu_{\delta}$}                 &
\colhead{Distance}                       &
\colhead{Separation}                     &
\colhead{$\theta$}                       &
\colhead{notes}                          \\

\colhead{}                               &
\colhead{(mas/yr)}                       &
\colhead{(mas/yr)}                       &
\colhead{(pc)}                           &
\colhead{}                               &
\colhead{(mas/yr)}                       &
\colhead{(mas/yr)}                       &
\colhead{(pc)}                           &
\colhead{($\arcsec$)}                    &
\colhead{($\degr$)}                      &
\colhead{}                               }

\startdata

 UPM 0111-7655A &   130.0 &    31.4 &    64.4 &  UPM 0111-7655B &   182.5 &    35.6 &   109.9 &  46.8 & 241.7 & \tablenotemark{a},\tablenotemark{f}                                        \\
 UPM 0245-8833A &   174.6 &    59.1 &    34.5 &  UPM 0245-8833B &   175.7 &    89.7 & \nodata &   6.8 &  15.0 & \tablenotemark{a}                                                          \\
 UPM 0320-4847A &   230.8 &   147.8 &    37.7 &  UPM 0320-4847B &   208.7 &    93.3 &   396.9 &  41.7 & 280.2 & \tablenotemark{a},\tablenotemark{b}                                        \\
 UPM 0533-5210A &   -80.5 &   180.7 &    58.2 &  UPM 0533-5210B & \nodata & \nodata & \nodata &   7.7 & 161.4 & \tablenotemark{a}                                                          \\
 UPM 0608-5301A &  -131.1 &   208.6 &    37.1 &  WD 0607-530B   &  -115.5 &   172.1 &   [98.8]&  21.5 & 120.7 & \tablenotemark{a},\tablenotemark{c},\tablenotemark{d}, 34.2pc              \\
 UPM 0740-5207A &   -37.7 &  -221.5 &    38.9 &  UPM 0740-5207B &     6.6 &  -240.0 & \nodata &  15.4 &  34.2 & \tablenotemark{a},\tablenotemark{e}                                        \\
 NLTT 19906A    &  -150.5 &   338.0 & \nodata &  NLTT 19907B    &  -150.5 &   338.0 & \nodata &   5.1 &  71.5 & \tablenotemark{f}                                                          \\
                &         &         &         &  UPM 0835-6018C &  -184.6 &   -17.7 &    50.3 & 113.0 &  49.3 & \tablenotemark{f}                                                          \\
\enddata

\tablenotetext{a}{Not detected during automated search but noticed by eye during the blinking process}
\tablenotetext{b}{Possible NLTT star with a position difference $>$ 90$\arcsec$ when compared to UCAC3 position}
\tablenotetext{c}{White dwarf, plate distance [in bracket] is incorrect with a more accurate distance in the notes}
\tablenotetext{d}{Number of relations used for distance estimate $<$ 6; plate distance less reliable}
\tablenotetext{e}{Source not in 2MASS}
\tablenotetext{f}{Proper motions suspect}
\tablenotetext{g}{SuperCOSMOS plate magnitudes suspect}

\end{deluxetable}


\end{document}